\documentclass[doublecol,linenumbers]{epl2} 
\usepackage{amssymb}
\usepackage{graphicx}
\usepackage{dcolumn}
\usepackage{bm}
\usepackage{longtable}
\usepackage{color}
\usepackage{pdflscape}
\usepackage{hyperref}

 \newcommand{\A}{\AA$^{-1}$} 
 \newcommand{\KHC} {KHCO$_3$} 

\newcommand{\KDP}{KH$_2$PO$_4$}

\newcommand{\Sch} {Schr\"{o}dinger}

\date{\today}

\title{A unified quantum-classical theory of the thermal properties of ice, liquid water and steam}

\author{Fran\c{c}ois Fillaux \inst{1}}
\shortauthor{F. Fillaux}

\institute{                    
  \inst{1} Sorbonne Université, CNRS, MONARIS, UMR 8233, F-75005, 4 place Jussieu, Paris, F-75005 France.}

\abstract{
The thermal properties of ice, liquid water and steam are at odds with statistical theories applied to many-body systems. Here, these properties are quantitatively explained with a bulk-scale matter field emerging from the indefinite status of the microscopic constituents. Such a field is characterized by its symmetry in spacetime, its degree of degeneracy and its eigenstates. There are several one-to-one correspondences bridging outcomes of classical and quantum measurements. (i) The heat capacities are linked to the symmetry of the field for each phase of water. (ii) The latent heats are linked to the change of the degree of degeneracy for each transition. (iii) The critical temperatures are linked to the eigenstates of the potential operator. The matter field leads to a complete representation of the phases of water, free of hidden parameters and statistical ignorance}

\begin{document}

\date{\today}

\maketitle

\section{Introduction}

Water, the matrix of life, covers the two-thirds of our planet. It is of central importance in physics and chemistry, earth and life sciences, cosmology and many technologies as well. No other material is commonly found as a solid, a liquid, or a gas at normal pressure. It is likely the most extensively studied material, but there is still no quantitative explanation of why its properties depart from statistical models concerned with an exponential number of molecules ruled by the classical equations of motion. Everyone is agreed that at least some of the ``abnormal'' properties stem from cooperative hydrogen bonds subject to strong local constraints referred to as the ``ice-rules''. Yet it is still not clear how this inter-linking translates into bulk-scale behavior and computer simulations are hardly conclusive, as they are notoriously sensitive to how the forces between molecules are modeled \cite{Ball,GY}. In fact, the large number of contending models indirectly indicates their lack of success in reproducing the properties of water and it is legitimate to investigate a decidedly new approach. 

Converging evidences of nuclear quantum effects in ice \cite{BKPS}, liquid water \cite{WCD,CDK} and steam \cite{KS}, demonstrate that the workings of these phases is quantum in nature, hence nonlocal, so a network of H-bonded molecules locally subject to the ice-rules is not a pertinent model \cite{Fil8}. Then, the question of what is the root level of reality for the quantum substrate of water and how thermal properties relate to this reality emerges. Obviously, this question challenges the divide between quantum and classical mechanics which led Bohr, and many others, to suppose an arbitrary limit in scale and complexity beyond which classical mechanics supersedes the quantum theory. However, this limit is not a consequence of the quantum theory in its own right, so different interpretations have been proposed but no decisive argument has prevailed so far. It is shown below that the quantum matter-field-in-a-box model quantitatively explains quantum and classical properties of water within a unified ``quantical'' framework without dividing boundary. 

\section{\label{sec:1}The quantum matter-field-in-a-box} 

A key point of a physical theory in materials science is the root level of reality it is concerned with, as compared with the continuous spacetime-translation symmetry (CSTTS) of the vacuum. In classical mechanics, the continuous symmetry is irreversibly broken. Classical objects have definite positions at any time in the space of everyday life, say the $x$-space. The classical view of water consists of an ensemble of nuclei ruled by the classical equations of motion. Ice is a dynamical lattice of O-atoms hosting an exponential number of proton configurations for which existence of molecular entities is a semi-subjective issue. Liquid water is a tetrahedral network of cooperative H-bonds in a jumble of molecular 
clusters which continually break and form. Steam is comprised of molecular dimers linked through fleeting H-bonds. Diversity, complexity and statistical ignorance hamper a formal description, or even a conclusive modeling, of the thermal properties. In addition, nuclear quantum effects are out of reach. Thermal properties are also outside the scope of quantum statistics concerned with an isolated ensemble of microstates out of equilibrium, as the unitary time-evolution hampers thermalization towards equilibrium \cite{Srednicki,Deutsch}. 

Contrary to classical mechanics, the reality underlying quantum mechanics is rather obscure, for the theory is designed purely to predict probabilities of outcomes of measurements. According to Bohr, this reality is basically unknowable. An individual entity viewed in isolation cannot be specified in terms of intrinsic characteristics in space and time prior to it being measured, and, when it is effectively measured, sometimes it exhibits a wave character, sometimes a particle character, and these states cannot be realized simultaneously (complementarity). 

Here, in order to avoid renunciation to objective knowledge, the root level of reality for an individual microscopic object in isolation is conceived of as a matter-field with a mass, an optional electric charge, and the continuous translation symmetry of the vacuum. Thus, space and time variables are indefinite prior to measurement. Wave-like and particle-like states are brought into being via measurements breaking the continuous symmetry in conformity with Bohr's complementarity. 

Pursuing the same line of reasoning, a bulk-scale matter-field emerges for a macroscopic ensemble of such quantum objects enclosed in a box. The mass is a constant and the center-of-mass obeys the classical equations of motion. 

At the microscopic level, the the vacuum state is the invariant ground state of the constituents which excludes the classical degrees of freedom (DOF). The constituents are outside the scope of mechanics and statistics, including thermodynamics. 

At the macroscopic level, the constituents subject to quantum degeneracy pressure behave collectively as a bulk-scale single-body whose number of DOF depends on the spacetime symmetry distinctive of the physical state of the system under consideration. In the case of water, the numbers are as follows. (i) Zero DOF for CSTTS. (ii) Breaking the continuous time-translation symmetry (CTTS) brings into being time and energy for 9 DOF$_\chi$ in a $\chi$-space separate from the $x$-space. Periodic time symmetry (PTS) brings into being a linear combination of bulk-scale eigenstates $|\Psi_n(\chi_i) \rangle$, $i = 1\cdots 9$ in conformity with \Sch's equation leading to equipartition of kinetic and potential energies. (iii) Breaking the continuous space-translation symmetry (CSTS) brings into being momentum in $\kappa$-space and conjugate positions relative to 9 DOF$_\xi$ in $\xi$-space of which the $x$-space is a subspace. As time is indefinite, lattice states with periodic space 
symmetry (PSS) are purely static and kinetic-energy is excluded. 
(iv) Breaking both symmetries brings into being $2\times9$ DOF in separate spaces. 

It follows that the molar heat capacity of water is a quantical property whose eigenvalues are $\frac{9}{2} \eta \mathcal{R}$. $\mathcal{R}$ is the gas constant and $\eta = \eta_\chi + \eta_\kappa$, with $\eta_\chi$ and $\eta_\kappa$ equal 0 for CTTS or CSTS, respectively, or 1 otherwise. 

\section{\label{sec:2} Quantical heat capacities} 

\begin{figure}
\begin{center}\includegraphics[angle=0.,scale=0.5]{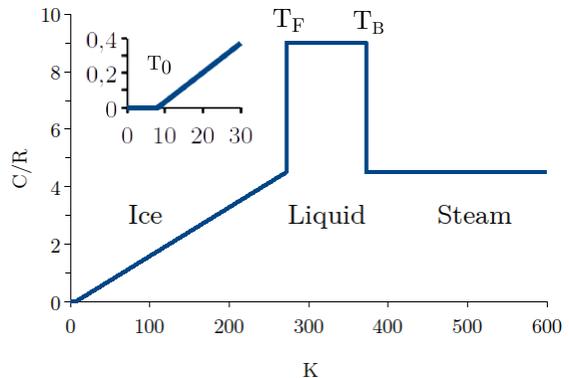}
\caption{\label{fig:1} The reduced specific heat capacity $C/\mathcal{R}$ of H$_2$O at normal pressure. }
\end{center}\end{figure}

An ideal measurement starts with a sample in diathermal equilibrium with a black-body at $T$. Diathermal means zero momentum transfer. An incremental jump $\delta T$ at time $t_0$ sets the system out-of-equilibrium. If there is no phase transition the amount of heat $\delta Q$ transferred reversibly until equilibration at $T + \delta T$ yields the heat capacity 
$C(T) = \lim (\delta T \rightarrow 0)\ \delta Q/\delta T. $
(From now on ``$C$'' stands for ``$C_P$'' at constant pressure.) If there is a phase transition, $\Delta Q$ is the latent heat. $C$ and $\Delta Q$ are classical quantities independent of the act of measurement and the precision of the measure depends on technical limitations exclusively. 

In many-body systems, the thermal energy is the sum of the mean kinetic and mean potential energy, which both equal $\frac{1}{2}k_B T$ per DOF. $k_B$ is Boltzmann's constant. Hence the molar heat-capacity of water is proportional to $9 \mathcal{R}$. The thermal energy, the population of the density-of-states (DOS) and the heat capacity diminish simultaneously with $T$ towards zero at 0 K. As $T \longrightarrow \infty$, the asymptotic limit is $9\mathcal{R}$. In between these limits, $C$ is an increasing function of $T$. 

The heat capacities of the phases of water (Fig. \ref{fig:1}) are clearly outside the scope of statistical physics. Instead, Table \ref{tab:1} substantiates the quantical one-to-one correspondence between $C$ and $\frac{9}{2}\eta\mathcal{R}$. (i) Below $T_0 \approx 8$ K, $C_{0H} = 0$. (ii) Above $T_0$, $C_{IH} = \frac{9}{2} \mathcal{R}$ for ice at the melting point $T_F$ or steam above the boiling point $T_B$ (the temperature law in ice is explained below). (iii) For liquid water $C_{WH} = 9\mathcal{R}$. There is no visible statistical effect due to a distribution of micro constituents in liquid water and steam. 

\section{\label{sec:3}Quantum measurements}

At the microscopic level, the continuous symmetry of the constituents can be broken by microevents consisting of (i) an input quantum ($h\nu_0, \mathbf{k}_0$), namely a photon or a neutron propagating in $x$-space, 
(ii) an output quantum ($h\nu_f, \mathbf{k}_f$) propagating in the same space, possibly lost in the environment or captured at a remote detector, (iii) an induced microstate ($h\nu = h\nu_0 - h\nu_f, \mathbf{k} = \mathbf{k}_0 - \mathbf{k}_f$) in $\chi$- or $\kappa$-space. The induced state and the output quantum are separate. The information conveyed by the output quantum concerns the induced state, exclusive of the system prior to the measurement. The induced microstate decays spontaneously to the ground state, so the field has no long term memory of past events. The density of microstates is negligible inasmuch as the incident flux is sufficiently low. 

\begin{center}
\begin{table} 
\caption{\label{tab:1} Reduced molar heat capacities and quantum numbers $\eta$ for the phases of H$_2$O. $\mathcal{R} \approx 8.314$ J.mol$^{-1}$.K$^{-1}$. }
\begin{tabular}{lllll}
& $T$ (K) & $2C/9\mathcal{R})$ & Ref. & $\eta_\chi + \eta_\kappa$ \\ 
\hline\\
Steam & $373.16 - 647.10$ & $1.001$ & \cite{Verma} & $1_\chi + 0_\kappa$ \\
Liquid & $273.16 - 373.16$ & $2.02$ & \cite{LMM} & $1_\chi + 1_\kappa$ \\
Ice Ih & $8-273.16$ & $1.01 \displaystyle{\frac{\Theta_{I}}{\Theta_{IF}}}$ & \cite{FW} & $1_\chi + 0_\kappa$ \\
Ice 0 & $0-8$ & $0$ & \cite{FW} &  $0_\chi + 0_\kappa$ \\
\end{tabular}
\end{table}
\end{center}

The case study of water is mainly concerned with three classes of complementary measurements. 

1) Diffractometers measure momentum transfer via elastic scattering. Each event realizes a snapshot with PSS and the statistics of outcomes give the momentum density in the $k$-space conjugate of the $x$-space, which are sub-spaces of $\kappa$- and $\xi$-space, respectively. Time is indefinite and there is no bound state. A complete representation in $\kappa$- or $\xi$-space is inaccessible. 

2) Photonic spectrometers (\textit{e.g.}, infrared, Raman, NMR...) measure energy transfer for $\mathbf{k} \approx 0$. The induced microstate with PTS is represented in the $\chi$-space and the eigenvectors are along the DOF$_\chi$ of the bulk. Dynamics and equipartition of kinetic and potential energy are invisible via quantum measurements conducted in $x$-space. 

3) Neutron spectrometers measure both energy and momentum transfer (\textit{e.g.}, inelastic neutron-scattering, aka INS, QENS, neutron Compton scattering, aka NCS...). They realize at random either $\mathbf{k}$-dependent states in $\chi$-space or static states in $\kappa$-space. 

\begin{center}
\begin{table} 
\caption{\label{tab:2} Thermal potential $\mathcal{E}$, thermal parameter $\Theta$ and degeneracy entropy $\mathcal{S}$ of the phases of H$_2$O. }
\begin{tabular}{llll|}
& $\mathcal{E}$ & $\Theta$ & $\mathcal{S} /\mathcal{R}$ \\
\hline\\
Steam & $\mathcal{R}T_B + \frac{9}{2}\mathcal{R}\Theta_{S}$ & $\Theta_{S} = T - T_B$ & $0$ \\
Liquid & $\frac{9}{2}\mathcal{R}\Theta_{IF} + 9\mathcal{R}\Theta_W$ & $\Theta_{W} = T - T_F$ & $ \ln14 + 2\ln\left(\displaystyle{\frac{3}{2}} \right) $ \\
Ice Ih & $\displaystyle{\frac{9}{2} \frac{\Theta_{I}^2}{\Theta_{IF}}}$ & $\Theta_{I} = T - T_{0}$ & $ \displaystyle{\frac{2\Theta_{I}}{\Theta_{IF}}} \mathcal{R}\ln \left(\frac{3}{2}\right)$ \\
Ice 0 & $0$ & $\Theta_{I0} = 0$ & $0$ \\
\end{tabular}
\end{table}
\end{center}

NCS data reported for ice and liquid water \cite{PSA,SRA} exemplify the complementarity of energy and momentum transfer. In statistical physics, this technique is supposed to probe a preexisting distribution of proton-momenta whose variance $\langle\sigma_\mathbf{k}^2 \rangle$ is linked to the mean kinetic-energy per proton via $\langle E_{K} \rangle = 3h^2 \langle \sigma_\mathbf{k}^2 \rangle /2m_H = 3k_B T/2$. Yet, $\langle E_{K} \rangle_{NCS}$ 
effectively measured is at variance with this law. In Fig. 4 in \cite{SRA} $\langle E_{K} \rangle_{NCS}$ is almost a constant for ice, namely $(154 \pm 4)$ meV at 5 K and $(162 \pm 4)$ meV at 271 K. The estimated slope, $ (0.02 \pm 0.02)$ meV.K$^{-1}$, is minute relative to $3k_B/2 \approx 0.12$ meV.K$^{-1}$. At $T_F$, $\langle E_{K} \rangle_{NCS}$ diminishes by $\approx 8\%$. 
This does not make sense in terms of kinetic energy. In the liquid, $\langle E_{K} \rangle_{NCS}$ increases quasi-linearly with $\Theta_W = T-T_F$, instead of $T$, and the slope is $\approx 0.1$ meV.mol$^{-1}$.K$^{-1}$. 

Contrary to statistical physics, the matter field excludes preexisting kinetic energy at the microscopic level. For ice, $\eta_\kappa =0$. Only elastic momentum transfer ($|\mathbf{k}| = 0$) is permitted. The statistics of outcomes gives a Gaussian profile analogous to the Debye-Waller factor $\exp - \mathbf{k}^2 \mathbf{.u}_{\chi I}^2$. The variance $\langle u_{\chi I}^2\rangle$ is the mean-square amplitude for proton displacements within the induced potential-operator and the weak temperature effect is representative of the anharmonicity. Kinetic energy is invisible. For the liquid, $\eta_\kappa =1$. Inelastic momentum-transfer is permitted. Such events realize the momentum distribution for ``free'' protons, namely $\frac{3}{2} k_B \Theta_W $, as there is no bound state in $\kappa$-space. This static distribution is not representative of a kinetic energy, as it is proportional to $\Theta_W$ instead of $T$. Uncorrelated elastic and inelastic events add up, so the variance effectively measured is $\langle \sigma_W^2 \rangle = \langle u_{\chi W}^2\rangle + \frac{3}{2} k_B \Theta_W $. At $T_F$, $\langle u_{\chi I}^2\rangle > \langle u_{\chi w}^2\rangle$ is in line with the lower density of ice relative to liquid water. 

\begin{figure}
\begin{center}\includegraphics[angle=0.,scale=0.5]{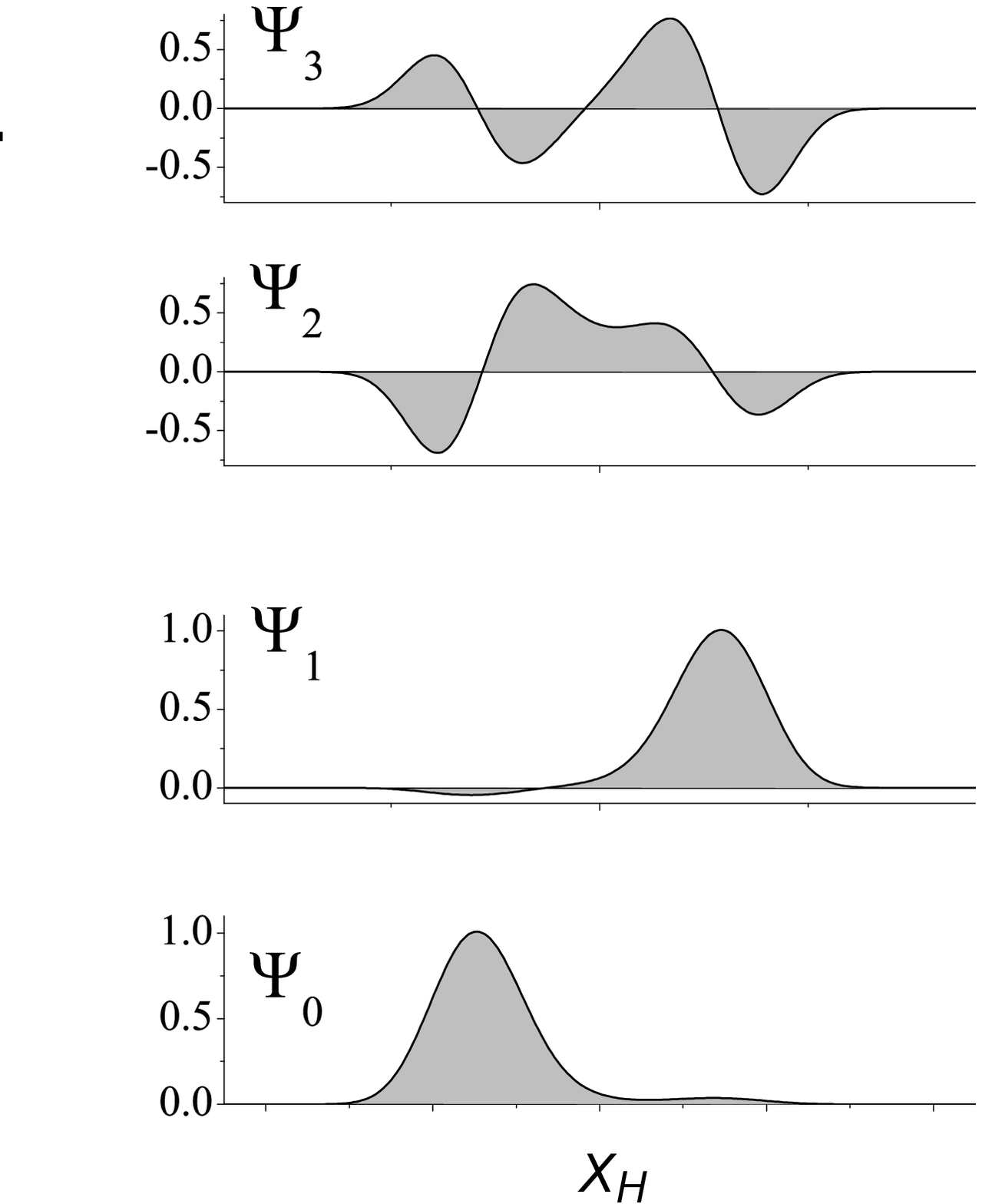}
\caption{\label{fig:2} Qualitative view of the eigenfunctions of the asymmetric double-well operator $\hat{v}(\chi_H)$. }
\end{center}\end{figure}

\section{\label{sec:4} Tunneling in ice}

In statistical physics, the key idea underpinning the ice-rules for a classical network of hydrogen bonds O$_d-$H$\cdots$O$_a$ between a donor and an acceptor groupings is that each proton experiences an asymmetric double-well potential, say $v(x_H)$, along a stretching coordinate $x_H$ in $x$-space. The asymmetry is the difference 
in energy between configurations O$_d-$H$\cdots$O$_a$ (say $L$) and O$_d\cdots$H$-$O$_a$ ($R$) corresponding to the lower and upper wells, respectively. The inter-well distance, $r$, the stretching frequency ($\nu$H), the barrier height at the bond center ($H$), the potential asymmetry ($\Delta V$) and the dipole moment ($\mu$) diminish as the O$_d\cdots$O$_a$ distance shortens, typically from $\approx 2.9$ \AA\ for weak H-bonds such as in ice, to less than $\approx 2.5$ \AA\ for strong and symmetric H-bonds \cite{Novak}. The shape of the potential is basically temperature-independent and symmetrization ($\Delta V =0$) at a constant pressure is impossible. 
Disorder in ice is commonly attributed to thermally-activated proton-jumps over the barrier, in conformity with the ice-rules. 

In quantum mechanics, the notions of chemical bonding vanishes. The proton coordinate turns into a nonlocal variable $\chi_H$ outside $x$-space. The eigenfunctions of the asymmetric double-well operator $\hat{v}(\chi_H)$ are sketched in Fig. \ref{fig:2}. The asymmetry $\Delta E = h\nu_1$ is the difference between the lower state in the upper well ($n = 1$) and that in the lower well ($n = 0$). The upper states are well separate ($h\nu_2$ and $h\nu_3 \sim 10\times h\nu_1$), so the two-level approximation is justified. For $E_R - E_L = h\nu_{1}$ the eigenstates of $\hat{v}_L$ read:
\begin{equation}\label{eq:1} \begin{array}{rcl}
|\psi_{L0}\rangle & = & \cos \theta |\psi_{0L}\rangle + \sin \theta |\psi_{0R}\rangle; \\
|\psi_{L1}\rangle & = & \sin \theta |\psi_{0L}\rangle - \cos \theta |\psi_{0R}\rangle; \\
\tan 2\theta & = & \displaystyle{\frac{\nu_t}{\nu_1 - \nu_t}}. \\
\end{array}\end{equation}
$\psi_{0L}$ and $\psi_{0R}$ are the zeroth-order wavefunctions localized within the $L$ or $R$ well, respectively. The mixing angle $\theta$ depends on the tunneling frequency $\nu_t$ extrapolated as $\nu_1 \longrightarrow 0$ at constant $r$ and $H$. Similarly, the eigenstates of $\hat{v}_R$ with opposite asymmetry $E_L - E_R = h\nu_{1}$ are: 
\begin{equation}\label{eq:2} \begin{array}{rcl}
|\psi_{R0}\rangle & = & \cos \theta |\psi_{0R}\rangle + \sin \theta |\psi_{0L}\rangle; \\
|\psi_{R1}\rangle & = & \sin \theta |\psi_{0R}\rangle - \cos \theta |\psi_{0L}\rangle. \\
\end{array}\end{equation}
For a dimer in isolation, $L$ and $R$ configurations are indistinguishable and superposition of (\ref{eq:1}) and (\ref{eq:2}) yields two tunneling splittings instead of only one for a genuine symmetric double-well \cite{FCou4}: 
\begin{equation}\label{eq:3} \begin{array}{rcl}
|\psi_{0\pm}\rangle & = & 2^{-1/2}[|\psi_{L0}\rangle \pm |\psi_{R0}\rangle]; \\
|\psi_{1\pm}\rangle & = & 2^{-1/2}[|\psi_{L1}\rangle \pm |\psi_{R1}\rangle]; \\
h\nu_{0\pm} & = & h\nu_{1\pm} = h\nu_t. \\
\end{array}\end{equation}

In the case of ice, the interpretation of neutron scattering data by Bove \etal \cite{BKPS} is not conclusive because it rests on the prejudice that tunneling splitting is excluded. Accordingly, these authors were concerned with a quasi-elastic Lorentzian profile of relatively low intensity underneath a huge elastic peak. However, they noted that the lack of temperature effects effectively observed is not in conformity with thermally activated jumps within an asymmetric double-well, so they concluded that quantum effects are dominant. 

With the matter-field, the induced tunneling splitting is dictated by the bulk-scale CSTS ($\eta_\kappa =0$). Visual inspection with fresh eyes of Figs 1 and 4 in \cite{BKPS} reveals two humps of intensity centered at $\approx\pm 0.1$ meV, respectively, so $h\nu_t/{k_B} = (1.2 \pm 0.2)\ \mathrm{K}$. This assignment is in line with the lack of temperature effect, since the bandwidth is representative of the instrument resolution ($15\ \mu$eV) and the intensity of the induced transitions is temperature independent. The profile of intensity as a function of $\mathbf{k}$ gives $\langle u_I^2 \rangle^{1/2} \approx 0.75$ \A and an effective oscillator mass of $ \approx  1$ amu. (It is shown below that $h\nu_t$ is consistent with calorimetry data.) 

Given that a genuine symmetric double-well is excluded for ice, if $\psi_{0\pm}$ were probability amplitudes subject to Born's rule, microevents would realize either $\psi_{L0}$ or $\psi_{R0}$ instead of the superposition. The observed tunneling splitting is an evidence in favor of the deterministic character of the wavefunctions leading to a non unitary radiative oscillation of the dipole moment. 

\section{\label{sec:5} The quantical phases of water}

\subsection{\label{sec:51} Ice Ih}

The heat capacity equals zero below $T_0 = (8 \pm 1)$ K and then increases quasi-linearly with $T$ above $T_0$ (Fig. \ref{fig:1} and Table \ref{tab:1}). Best fit exercises with polynomial functions inspired by Debye's model require 13 parameters in the range $0.5 - 38$ K and additional parameters above 30 K \cite{SLL,MK}. Yet, there is no obvious justification. 

Contrary to Debye's model, the discontinuity at $T_0$ demonstrates a phase transition of the second order and $\mathcal{R} T_{0} = 7\mathcal{N}_Ah\nu_t/k_B = (8.4 \pm 1.4)$ K is the tunneling gap for 7 molar operators $\hat{V}(\chi_{j_H}) = \mathcal{N}_A\hat{v}(\chi_{j_H}) $, $j = 1 \cdots 7$, in a space expanded into $7\times 1$-D$_{\chi_H}$. $\mathcal{N}_A$ is Avogadro's constant. (The hepta dimension is confirmed below by the heat of fusion.) $\mathcal{R} T_{0}$ is the energy necessary to break the CTTS blocking heat transfer. This state is independent of the act of measurement, so the wavefunction is deterministic. The gap is dictated by the indefinite status of the constituents. It demonstrates tunneling of a single-body, without internal DOF in hepta dimensions, whose mass is $\approx 7$ g.Mol$^{-1}$. Bohr's divide is ruled out.

The antisymmetric state at $\mathcal{R} T_0$ reads $\mathcal{N}_A^{1/2} \bigotimes_{j = 1}^7 |\psi_{0-} (\chi_{j_H}) \rangle$ and the symmetric state $\mathcal{N}_A^{1/2} \bigotimes_{j = 1}^7 |\psi_{0+} (\chi_{j_H}) \rangle$ at zero-energy is forbidden by $\eta_0 = 0$. The linear variation of $C_{IH}$ in the range $T_{0} - T_{F}$ reads: 
\begin{equation}\begin{array}{l}\label{eq:4} 
|\Psi_{I} \rangle = \mathcal{N}_A^{1/2} \bigotimes\limits _{j = 1}^7 [|\psi_{0-} (\chi_{j_H}) \rangle \\
+ \alpha_{I}(|\psi_{1+} (\chi_{j_H}) \rangle + |\psi_{1-} (\chi_{j_H}) \rangle]. \\
\end{array}\end{equation}
As a consequence of the dipolar coupling, $\alpha_{I} = \Theta_{I}/{\Theta_{IF}}$ is real and $\Theta_{I} = T-T_0$ is the temperature in $\chi$-space which has no counterpart in $x$-space. 

The time-dependent part of $|\Psi_{I}|^2$, 
\begin{equation}\label{eq:5} \begin{array}{l}
|\Psi_{I}(t)|^2 - |\Psi_{I}(0)|^2 = \\ 
2\mathcal{N}_A \alpha_{I} \sum\limits_{j = 1}^7 \{ |\psi_{0-}(\chi_{j_H})||\psi_{1+}(\chi_{j_H})| \cos 2\pi(\nu_{1} -\nu_t/2) t\\
+ \alpha_{I}^2 |\psi_{1+}(\chi_{j_H})||\psi_{1-}(\chi_{j_H})| \cos 2\pi\nu_t  t \},\\
\end{array}
\end{equation} 
describes classical oscillations of the bulk-scale dipole moment and the time-average radiated power is 
\begin{equation}\label{eq:6} 
\mathcal{P} \approx \displaystyle {\frac {7\alpha_{I}^2 |M|^{2}}{3\pi\epsilon_0 c^3}} \{[2\pi(\nu_{1} -\nu_t/2)]^{4} + \alpha_I^2[2\pi\nu_t]^4\}.
\end{equation}
$\epsilon_0$ is the permittivity of free space, $c$ is the speed of light and $|M|$ is the amplitude of the molar dipole moment. Since the radiance at $h\nu_t$, relative to that at $h(\nu_1 -\nu_t/2)$, is minute for a black body, $C = d\mathcal{P} /d\Theta_{I}$ is practically proportional to $\Theta_{I}$. Apart from the $\chi_{jH}$'s, the contribution of the other DOF$_\chi$'s is invisible. 

The thermal energy of ice in isolation prepared at $T > T_0$ at a given instant of time decreases spontaneously and vanishes at $T = T_0$ (Table \ref{tab:2}). Otherwise, ice in contact with a black-body at $T > T_0$ is stationary when the net flux of input and output photons equals zero. After a $T$-jump, photons matching the beat frequencies contribute to heat transfer to the $\chi_{jH}$'s and the power radiated yields equipartition among the 9 DOF$_\chi$. There no time reversal symmetry and the ``eigenstate thermalization hypothesis'' \cite{Srednicki,Deutsch} is pointless. 

\subsection{\label{sec:52} Liquid water}

Fusion breaks the CSTS and separates the ice state in $7\times 1$-D$_\chi$ into 14 degenerate states. This is attested by the heat of fusion, given that the entropy of statistical physics is replaced with the degeneracy entropy of the field. Let $W_{WF}$ and $W_{IF}$ be the degrees of degeneracy for ice and liquid water, respectively, and $\Delta W_{F} = W_{WF} - W_{IF}$. Then, the heat of fusion $\Delta H_{FH} \approx 6007\ \mathrm{J.mol}^{-1}$ of ordinary ice \cite{FW} yields:
\begin{equation}\begin{array}{rcl}\label{eq:7} 
\Delta W _{FH} = \displaystyle{\exp\frac{\Delta H_{FH}}{\mathcal{R}T_{FH}}} & \approx & 14.08.\\
\end{array}\end{equation}
For D$_2$O ice, $\Delta W_{FD} \approx 14.34$ is comparable \cite{Fil8}. 

\begin{figure}
\begin{center}\includegraphics[angle=0.,scale=0.3]{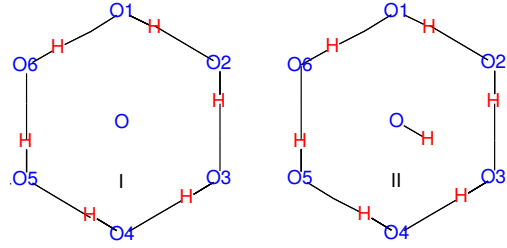}
\caption{\label{fig:3} Symbolic ``clockwise'' graphs for liquid water. }
\end{center}\end{figure}

The 14 states can be translated into symbolic graphs of nonlocal pairwise correlations. In Fig. \ref{fig:3}, each graph is comprised of a vertex ``O'' at the center of a ``clockwise'' hexagonal buckle of distinguishable vertexes $\mathrm{O1, \cdots O6}$, with oriented ``OHO'' edges. \textbf{I} is single and \textbf{II} is one among six equivalent graphs. Fusion demonstrates separation of the ice state into 7 clockwise and 7 anticlockwise states. In addition, each H-vertex of the hexagonal buckle is correlated with 4 indistinguishable O-edges and $1/4$ H-vertex per O-edge is correlated with 6 distinguishable O-vertexes, so the degree of degeneracy is $3/2$. Similarly, the degeneracy of the ``interstitial'' H-vertex equals $3/2$, so the total degree of degeneracy is $W_W = 14(3/2)^2$. The heat capacity testifies that $W_W$ is a constant in the range $T_{FH} - T_{BH}$ (Table \ref{tab:2} and Fig. \ref{fig:5}). 

\begin{figure}
\begin{center}\includegraphics[angle=0.,scale=0.35]{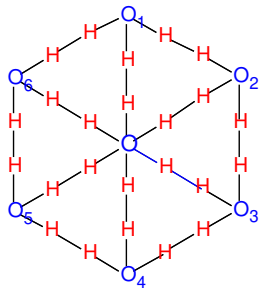}
\caption{\label{fig:4} Symbolic graph for ice Ih. }
\end{center}\end{figure}

For ice at $T_F$, $W_{IF} = \left(3/2\right)^2$ translates into a superposition of 14 graphs of distinguishable vertexes (Fig. \ref{fig:4}). In the range $T_F-T_0$ the degeneracy entropy is proportional to $\Theta_{IH}$ and equals zero at $T\le T_0$ (Table \ref{tab:2} and Fig. \ref{fig:5}). 

\begin{figure}
\begin{center}\includegraphics[angle=0.,scale=0.5]{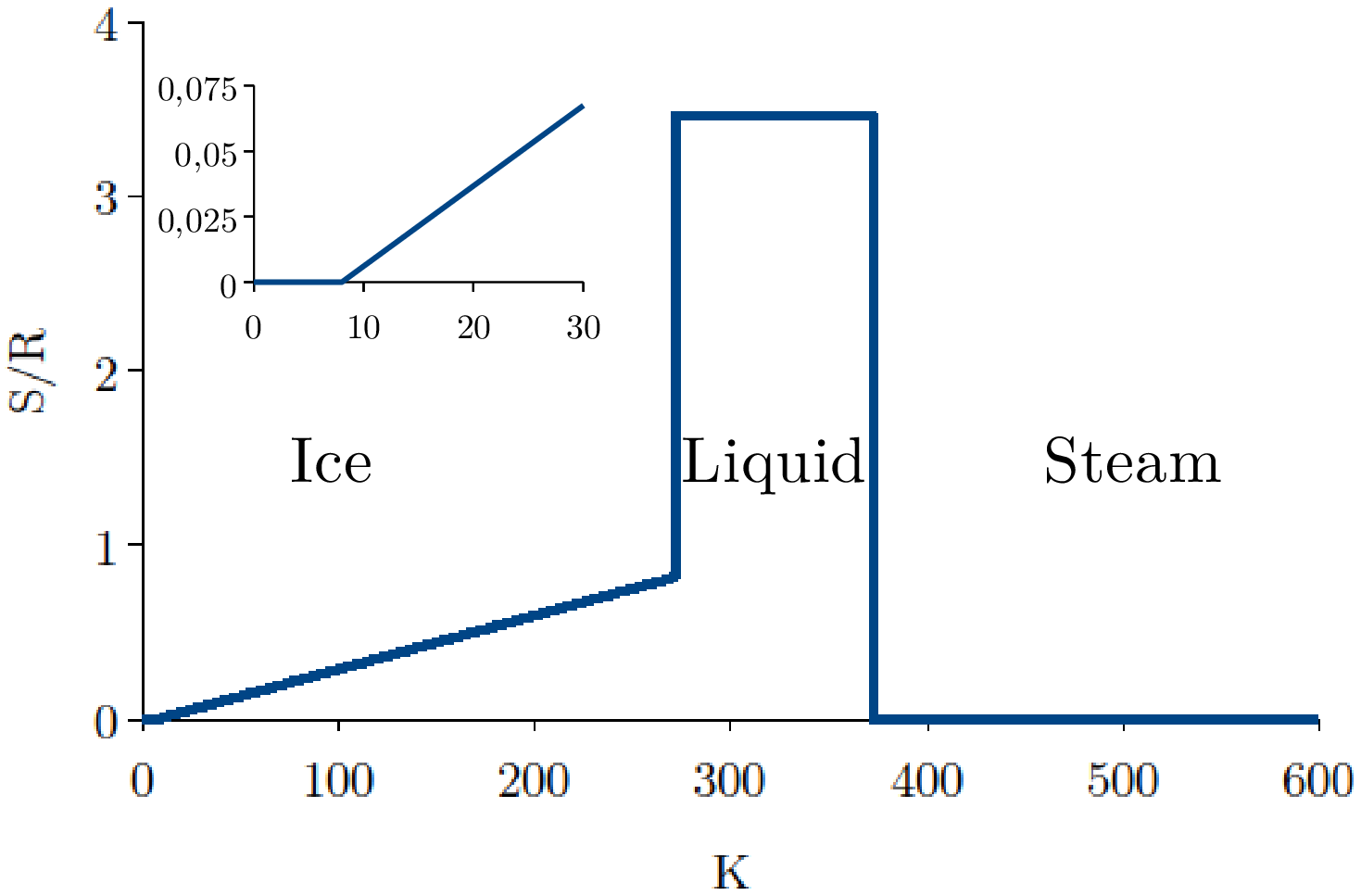}
\caption{\label{fig:5} The reduced degeneracy entropy $\mathcal{S}/\mathcal{R}$ of H$_2$O. }
\end{center}\end{figure}

$T_F$ and $T_B$ are linked to the eigenstates of $\hat{V}$ as follows. Continuation of (\ref{eq:4}) above $T_F$ would involve higher eigenstates whose delocalized wavefunctions $\psi_2$ and $\psi_3$ (Fig. \ref{fig:2}) are antibonding, so the field would be destroyed. Instead, $\hat{V}$ splits into $\hat{V}_L$ and $\hat{V}_R$ whose eigenstates are shifted upwards, so the antibonding states are avoided. At $T_F$ the thermal energy of ice (Table \ref{tab:2}) equals the potential energy, 
\begin{equation}\label{eq:8} 
\displaystyle{\frac{9}{2} \mathcal{R}\Theta_{IF}} = \displaystyle{\frac{7}{2} \mathcal{N}_A [h(\nu_{1I}+\frac{1}{2}\nu_t) + h(\nu_{1I}-\frac{1}{2}\nu_t)],} 
\end{equation}
so $h\nu_{1I}/k_B \approx 170\ \mathrm{K}$. At $T_B$, the same reasoning applied to $\hat{V}_L$ and $\hat{V}_R$ gives $9\mathcal{R}T_B = 14 \mathcal{N}_A h\nu_{1W}$ and $h\nu_{1W} / k_B \approx 240$ K. From NCS, $h\nu_{1I} <  h\nu_{1W}$ correlates with a shorter inter well separation in water relative to ice. 

Heat transfer to liquid water is controlled by $h\nu_{1W}$, so the heat capacity is temperature independent. $C_W$ indicates equipartition among $\chi$'s and $\kappa$'s DOF, but equipartition of kinetic and potential energy is permitted in $\chi$-space, whereas only potential energy is permitted in $\kappa$-space. 

Spectroscopic measurements of $h\nu_{1I}$ and $h\nu_{1W}$ are lacking. However, $h\nu_{1W}$ makes sense as compared with INS transitions reported for similar asymmetric systems: $\approx 300$ K for potassium hydrogen carbonate (\KHC) \cite{FTP} and potassium dihydrogen phosphate (\KDP)\cite{FCou4} or $\approx 260$ K for benzoic acid \cite{FCou5}. 

\subsection{\label{sec:53} Steam}

At $T_B$, the energy available for vaporization includes the latent heat effectively transferred to the liquid, $\Delta H_{B} \approx 40657\ \mathrm{J.mol}^{-1}$ \cite{Marsh}, plus the thermal energy, plus the energy of the degeneracy entropy (Table \ref{tab:2}): 
\begin{equation}\begin{array}{rcl}\label{eq:10} 
\Delta E_{B} & = & \Delta H_{B} + \mathcal{R} T_{B} \left[ 9 + \ln 14 + 2\ln \frac{3}{2} \right ] ;\\
& \approx & 79266\ \mathrm{J.mol}^{-1}. \\ 
\end{array}\end{equation}
Comparison with the dissociation energy of dimers (H$_2$O)$_2$ in molecular jets, namely $D_{0} = (13.2 \pm 0.12)10^{3}\ \mathrm{J.mol}^{-1}$ \cite{RCM}, gives $\Delta E_{B} = (6.00 \pm 0.05) D_{0} $. Accordingly, vaporization converts 14 heptamers composed of 7 distinguishable dimers into a single dimer ($W_S =1$). The heat capacity of steam testifies that $W_S$ is temperature independent. 

Experimental data indicate that the potential operator of steam is quite similar to that of ice, yet shifted upwards by $\mathcal{R}T_B$. 
Firstly, the tunneling splitting of $\approx 0.94$ K for dimers in molecular jets \cite{OHP} compares with $h\nu_t$ in ice. Secondly, the critical temperature of steam, $T_{c} \approx 647.1$ K \cite{Verma}, relative to the boiling point, that is $T_{c} - T_{B} \approx 273.94 \ \mathrm{K}$, is similar to $T_{F}$. Thus, replacing $\Theta_{IF}$ in (\ref{eq:8}) with $(T_{c} - T_{B})$ yields $h\nu_{1S} / k_B \approx 176$ K. It follows that the hepta dimension of the $\mathcal{H}$-field is retained in steam, so the symbolic graph is the same as that for ice, except that the vertexes are indistinguishable. 

At $T_{B}$, the tunneling gap $\pm\mathcal{N}_Ah\nu_t / {2}$ is realized at no energy cost, so there is no threshold temperature for heat transfer, analogous to $T_0$ for ice. Steam is a radiative vapor whose heat capacity is temperature independent. 

The transition to the supercritical phase is analogous to the melting of ice. At $T_{c}$, the ice-like dimer splits into water-like dimers ($W_{Sc}= 2)$) and the second order character of the transition indicates a thermal equilibrium above $T_{c}$. 

\begin{center}
\begin{table} 
\caption{\label{tab:3} Correspondence between the critical temperatures and the eigenstates of the $\mathcal{H}$-operator: $h\nu_t$; $h\nu_{1I}$ and $h\nu_{1W}$. }
\begin{tabular}{l}
$T_0 = 7 \displaystyle{\frac{h\nu_t}{k_B}}$  $T_F = T_0 + \displaystyle{ \frac{7h\nu_{1I}}{9k_B}}$  $T_B = \displaystyle{ \frac{7h\nu_{1W}}{9k_B}}$  $T_c = T_B + \displaystyle{ \frac{7h\nu_{1I}}{9k_B}}$\\
\end{tabular}
\end{table}
\end{center}

\section{\label{sec:6} Remarks}

The time-dependent crystal of ice is different in nature from Wilczek's time crystal \cite{Wilczek2} and Floquet crystals \cite{SZ}, which are concerned with nuclear motions. 

The matter field is at variance with the quantum-lattice-gauge theory (QLGT) of ice proposed by Benton \etal \cite{BSS2}. These authors argue that concerted proton tunneling in hexagonal rings yields a coherent superposition of an exponentially large number of preexisting microstates, such that protons continually fluctuate between different configurations in conformity with the ice-rules. This view is clearly at variance with $C_0 = 0$ below $T_0$, which excludes fluctuations, and with the gap $\mathcal{R}T_0$, which excludes preexisting microstates. In addition, the QLGT introduces a locally causal relationship between tunneling and proton 
displacements in $x$-space that is excluded by Bell's theorems. 

Figs \ref{fig:3} and \ref{fig:4} are in line with interstitial molecules revealed by diffraction studies of ice and liquid water \cite{Soper,NL}. 

Freezing-point-depression studies of HCl-water solutions are in line with the hepta dimension of the field. These studies reveal stoichiometric heptamers H$^+$(H$_2$O)$_n$, $n = {6.7\pm0.7}$ whose binding energy of $(54 \pm 6)\times 10^3$ J.mol$^{-1}$ ensures stability up to the boiling point \cite{Zavitsas}. 

The degeneracy entropy of ice is different in nature from Pauling's ``residual entropy'' \cite{Pauling2}. Pauling was concerned with six-member rings without interstitial molecules leading to $\left(3/2\right)^{\mathcal{N}_A}$ degenerate configurations per mole in accordance with the ice-rules. He concluded that the residual-entropy $\mathcal{R}\ln (3/2)$ is temperature independent, yet with a serious caveat: ``Measurements made under ideal conditions and extended to sufficiently low temperatures would presumably lead to zero entropy for any system.'' This is effectively the case for the degeneracy entropy. Actually, Pauling's entropy compares with the mean value of $\mathcal{S}_I(\Theta_{I})$ in the range $T_0-T_F$ (Table \ref{tab:2}). 

The links between the eigenstates of the $\mathcal{H}$-operator and the critical temperatures (Table \ref{tab:3}) show that calorimetry and spectroscopy measurements are redundant. These links rule out the quantum-classical divide. 

All together, Tables \ref{tab:1}-\ref{tab:3} give a complete, self-contained and deterministic description of the thermal properties of water, substantiated with experimental data exclusively. This is likely the most parsimonious and accurate representation ever published showing that the indefinite status of the constituents and the phase-dependent bulk-scale symmetry are both necessary and sufficient elements of reality. 

Calorimetry is unique to study quantical properties, but comprehensive studies are rare. For the time being, it is unclear whether the matter field is of general significance for complex systems, although there is no foreseeable principled limitation. 

\section{\label{sec:7} Conclusion}

The quantical theory quantitatively explains quantum and classical properties of the phases of water and their transitions. The bulk-scale matter-field emerges from the indefinite status of the constituents and quantical physics applies when the continuous time-translation symmetry is broken. Quantical physics is ruled by \Sch 's equation in an abstract space outside the classical world of everyday life. The eigenfunctions are deterministic and the time evolution is not unitary. Quantical physics is in conformity with complementarity and nonlocality, yet it is at odds with Born's rule, \Sch 's cat thought experiment and decoherence. Furthermore, quantical physics is outside the area of classical mechanics and statistics, including thermodynamics. 

Determinism and nonlocality lead to a self-contained description of the phases of water substantiated with several one-to-one correspondences between classical and quantum features. (i) The heat capacities are linked to the symmetry of the quantum field. (ii) The latent heats are linked to the degree of degeneracy of the field. (iii) The critical temperatures are linked to the potential operator. The matter field explains why ice below $T_0$ is a thermal insulator in conformity with the definition of an ideal classical object without any internal degree-of-freedom. Ice above $T_0$ demonstrates a bulk-scale superposition state. Ice and liquid water are bulk-scale heptamer states. Steam and the supercritical fluid are bulk-scale dimer states. Heat transfer is driven by oscillations of the bulk-scale dipole moment and the internal temperature scale is phase-dependent. 

The case study of water highlights the quantical nature of our environment and, perhaps, of the live matter. 

\bibliographystyle{eplbib}

\end{document}